# When is a DAO Decentralized?

Henrik Axelsen*, Johannes Rude Jensen, and Omri Ross

Department of Computer Science, University of Copenhagen, Universitetsparken 5,
DK-2100 Copenhagen, Denmark

heax@di.ku.dk, johannesrudejensen@gmail.com, Omri@di.ku.dk

**Abstract.** While previously a nascent theoretical construct, decentralized autonomous organizations (DAO) have grown rapidly in recent years. DAOs typically emerge around the management of decentralized financial applications (DeFi) and thus benefit from the rapid growth of innovation in this sector. In response, global regulators increasingly voice the intent to regulate these activities. This may impose an excessive compliance burden on DAOs, unless they are deemed sufficiently decentralized to be regulated. Yet, decentralization is an abstract concept with scarce legal precedence. We investigate dimensions of decentralization through thematic analysis, combining extant literature with a series of expert interviews. We propose a definition of "sufficient decentralization" and present a general framework for the assessment of decentralization. We derive five dimensions for the assessment of decentralization in DAOs: Token-weighted voting, Infrastructure, Governance, Escalation and Reputation (TIGER). We present a discretionary sample application of the framework and five propositions on the future regulation and supervision of DAOs. We contribute new practical insights on the topic of compliance and decentralized organizations to the growing discourse on the application of blockchain technology in information systems (IS) and management disciplines.

## 1  Introduction

In financial markets, regulatory objectives traditionally focus on (1) proper functioning and integrity of markets, (2) financial stability, (3) protecting the collective interests of consumers and investor protection, while also (4) aiming to reduce criminal activity and (5) preserving monetary sovereignty.

The crypto economy has experienced rapid growth in recent years, amounting to USD 3 Trillion in late 2021 [1]. Due to its open-source nature, the sector is subject to high competition and enables decentralized finance (DeFi). DeFi replicates traditional financial services; hence the industry is becoming increasingly important to regulators [2], [3].

---

* Corresponding author

The crypto economy operates on permissionless blockchain technology. Regulators see this technology as imperative to innovation, growth, and global competitiveness. While crypto remains primarily unregulated, regulators across the globe are motivating and implementing crypto regulation to meet the challenge of ensuring consumer protection, innovation, and growth without stifling innovation [4], [5].

In recent years, scholars from a wide variety of disciplines have found a shared interest in examining the implications of the technical properties of blockchain technology in their fields. Concepts such as the self-enforcement and formalization of rules, automatization, decentralization of authority, transparent execution of business processes, and codification of trust appear to be conducive to wide-ranging theoretical and industrial innovation.

While there are multiple working definitions of the concept of decentralized autonomous organization (DAO) in industry, most take the form of fluid organizations or loosely organized communities, self-directed and governed through smart contracts without the presence of central authority or a managerial hierarchy [6], [7].

DAO tends to operate through bottom-up interaction and coordination among a set of independent and distributed rational agents. This has increased interest in how DAOs can mitigate principal-agent problems and reduce misconduct by improving [8] through shifting power dynamics. Some observers compare DAOs to nation-states rather than traditional organizations [9]. In this analogy, the formal (on-chain) smart contracts are comparable to a "computational constitution." At the same time, cultures are nurtured through communication emerging around the design, development, and maintenance of the products governed by the DAO.

While Ethereum remains the dominating network, DAOs are now proliferating across blockchains, facilitated by innovation in the underlying infrastructure. There are currently some 5 000 individual DAOs, counting more than 1.7m token holders, and some 700 000 active voting members [10].

Implementing regulatory objectives imposes a high compliance burden for industry participants [14] in traditional finance. For European actors, the total cost of compliance ranges between 2 and 25% of total operating expenses, depending on the size and complexity of the institution [11], [12]. Being subjected to traditional financial institutions' comparatively strict compliance requirements may prove challenging, if not impossible, for DAOs as they are designed today. Regulatory compliance imposes capital and liquidity requirements, strong centralized controls and separation of functions, management hierarchies, and complicated reporting.

Hence, if existing regulation is applied without scrutiny, the novel and poorly defined concept of a DAO may give rise to both conventional and emerging regulatory risks. A key driver among these risks is the prevailing ideological assumption that for regulation to have an effect, a subject in the form of a legal or physical person is required to be held accountable for obligations arising from DAO activities, including those related to regulated financial activities.

Recently, global regulators indicated that the issuance of crypto assets, which may otherwise be subject to compliance requirements, may be exempt if distributed by an entity predominantly or exclusively operating as a "decentralized entity" [5], [13].

Yet, none of the proposals published to date offer a working definition of what might constitute "sufficient decentralization."

As follows, designing a decentralized crypto-based business model based on "smart contracts" is complicated: In addition to the usual challenges in finding product market fit, product leadership, sales, recruitment, development, and scaling, founders must seek to operate their projected business in a decentralized manner or risk negative regulatory implications [14].

While founders may opt for the "Nakamoto model" [15] and operate in full anonymity, secondary service providers required to fund and execute a project are also subject to regulation. Consequently, fully anonymous (anon) stakeholders may find themselves operating in a vacuum, with limited access to ancillary services.

This article asks the following research question: *"When is a DAO (sufficiently) decentralized?"* We present an artifact designed to assess the level of decentralization in any given DAO across several dimensions. We seek to contribute new practical and actionable insights on the topic of decentralized organizations to the growing distributed ledger technology (DLT) discourse in the information systems and management disciplines. Further, we contribute to the growing regulatory discourse in crypto assets and decentralized finance by providing a pragmatic assessment tool for regulatory compliance assessment.

## 2 Background

**2.1 Blockchain Technology and "Decentralized Autonomous Organizations"**

Blockchain is a subset of DLT where transactions are recorded through immutable cryptographic signatures. A blockchain's primary function is maintaining an append-only ledger in a peer-to-peer network [16], using a consensus mechanism to validate transactions. Permissionless blockchains are decentralized computer networks that maintain a single global version of a shared database and a shared account ledger that is visible to all stakeholders [17]. Permissionless blockchains are open, so anyone can join, leave, read, and write as they please. No central party authorizes access, and its cryptographic primitives ensure collusion resistance [18]. Bitcoin [15] and Ethereum [19] are important instances of permissionless blockchains.

DeFi apps are financial solutions built with "smart contracts" operating through permissionless blockchain technology.

Smart contracts are scripts that automatically carry out specific business logic. Financial services or products created as smart contracts work autonomously without the need for monitoring or intervention from the software developers who originally designed the application due to the deterministic characteristics of the underlying blockchain.

This means that, as long as the blockchain is active, a smart contract will execute business logic unconditionally and irreversibly [20]. Typically, a smart contract will carry out a set of instructions that allow participants to lend or swap an underlying base asset or other financial assets that have been "tokenized" [21]. DAOs utilize these properties to create rules-based organizations, in which they make decisions instituted in code. A DAO will typically consist of multiple interacting smart contracts responsible for different parts of the DAO, including treasury management, the tallying of votes, and the token itself. All these smart contracts are deployed on the blockchain and maintained as stateful applications. Both users and smart contracts are represented by addresses and compute transactions in the database containing instructions on how to change the state. Transactions emitted to the network are then sequenced in blocks and circulated with the network, at which point a global state-change is enacted.

To illustrate the above, in Figure 1 we present a layered taxonomy in which the *protocol layer* represents the consensus model determining the logic by which blocks are generated and distributed; the *application layer* represents the virtual machine in which smart contracts are deployed, and the *interface* and *user* layers represent the web-based interface through which users can create and sign transactions.

When a user participates in DAO voting, this process is carried out through one or more transactions in which the user (1) maintains a balance of governance tokens on an address to which they control the private keys and (2) connects their wallet to sign a message or a transaction, enabling them to signal their approval or dismissal of a governance proposal.

While there are multiple ways to implement this logic, the leading solutions rely either on the collection of off-chain signatures through a voting interface (User A) or the direct collection of votes and implementation of pre-deployed code changes by the DAO contract (User B).

In response to voter apathy, DAOs may implement the option for vote-delegation. This is typically carried out directly in the token contract and implemented as a feature in which a token holder can assign the voting power associated with their balance to a third-party address without losing custody of the tokens.

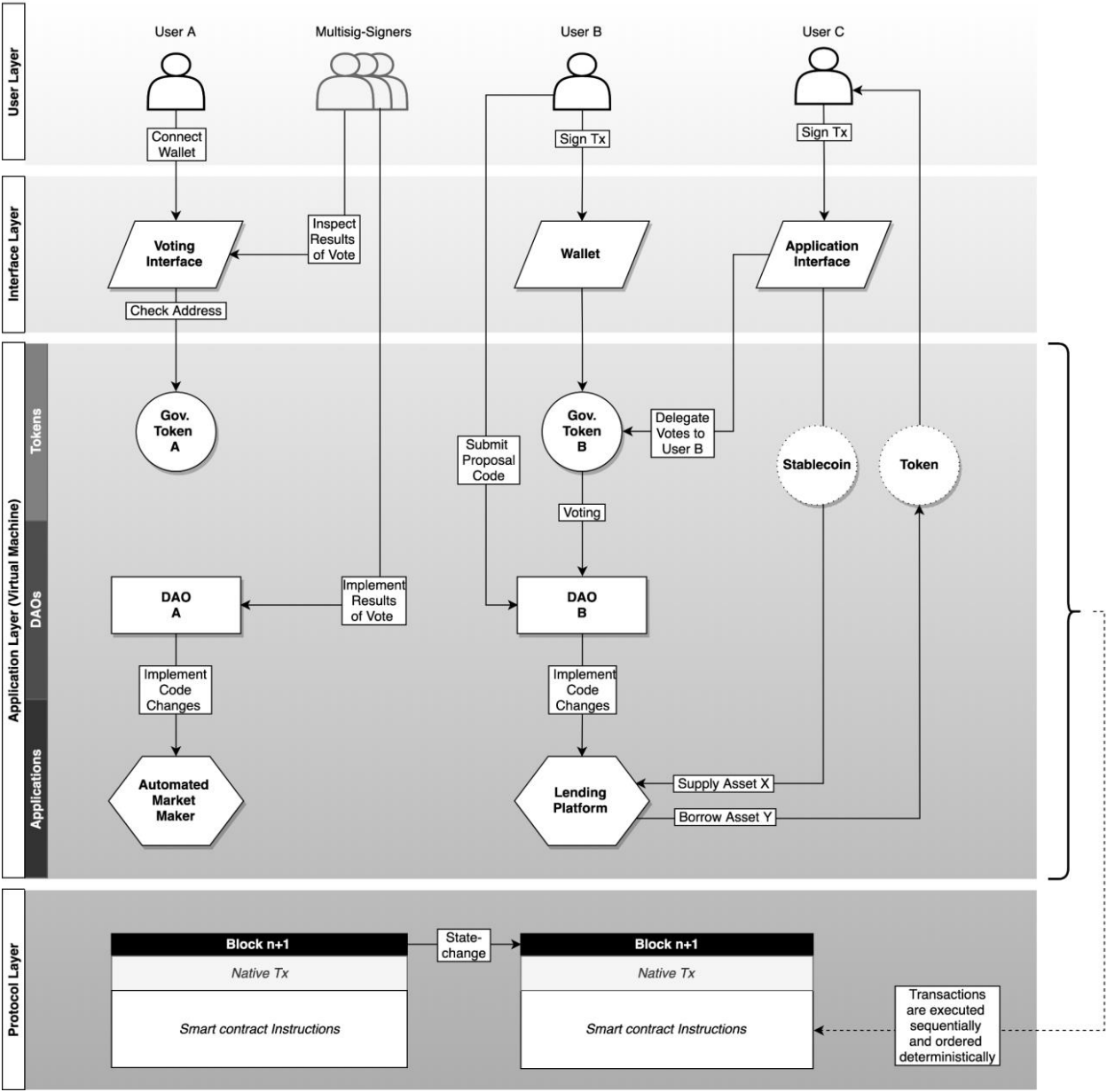

**Figure 1.** Blockchain, application layer, and users

**2.2 The Problem of Defining Decentralization within a Regulatory Context**

DAOs are mostly designed and instantiated by a small group of individuals who distribute power and control governance, with a promise to decentralize the governance process at some defined later stage [22].

Without legal recognition, most jurisdictions today may simply treat unregistered DAOs as unincorporated general partnerships, resulting in community members having personal, joint, and several liability for debts or legal actions arising from operating the DAO.

Increasingly, therefore, DAOs establish themselves with "legal wrappers" to protect DAO participants from unlimited liability, optimize tax treatment or engage in contractual "off-chain" transactions, even if not focused on regulatory compliance expectations and "sufficient decentralization" [23].

Because the common instantiation method is centralized from a design perspective, such a "wrapper" constitutes incorporation. It relates only to the autonomy and legal capacity of the organization, which technically does not prevent the concept of decentralization. Yet, DAOs that operate using a governance token, issued with a "*reasonable expectation of profits to be derived from the entrepreneurial efforts of others,*" are likely to be considered to undertake regulated financial activity [13].

Some scholars propose that a DAO, like autonomy classification for land and maritime environments [24], be considered autonomous to the extent that it can legally accept liability [22]. In practice, the level of autonomy and anonymity can vary, but a DAO is normally self-directed through voting on- and off-chain; it can be financial or non-financial in purpose, but the traditional legal system seems secondary to its existence and purpose [25].

In 2018, a US Securities and Exchange Commission (SEC) representative suggested that contractual and technical ways exist to structure digital assets, so they function more like consumer items or community enablers and less as regulated securities. At the same time, it was suggested that a security could become "sufficiently decentralized" over time so that it no longer is a security token under the so-called Howey test [13]. Since then, likely accelerated by the increasing success of DeFi, regulators across the globe have increasingly looked to regulate DeFi and DAOs, and uncertainty has prevailed.

Efforts to regulate DAOs as limited liability companies have emerged [26], [27]. More recently, progressive senators in the US are working on regional regulation of DAOs, yet this is still early draft, subject to extensive negotiations of political views [4].

As the first major region attempting to regulate crypto assets at the supranational level, the EU bloc emerged in 2020 with a digital finance package. The EU draft regulation included DAOs in the negotiation phase [5] with legal identity and limited liability for the community members. However, it was omitted in the final version of the regulation, called the Markets in Crypto Asset (MiCA) regulation, approved on June 30, 2022.

Much remains to clarify how DAOs will eventually become regulated, likely through a global policy setter, given the nature of DLT and the world-wide-web. At the time of writing, the final MiCA text is not published. Still, based on the EU Council's negotiation mandate, the regulation appears to treat decentralized activity in a manner similar to the US: "*This regulation applies to natural and legal persons and the activities and services performed, provided or controlled in any manner, directly or indirectly, by them, including when part of such activity or services is performed in a decentralized way…Where crypto assets have no offer or and are not traded in a trading platform which is considered to be operated by a service provider, the provisions of (this regulation, ed.) do not apply*" [28] (recital 12a).

This EU regulation appears to align with the global trend that certain crypto assets may become exempt from specific compliance requirements, even if constituting an activity that might otherwise be a regulated financial activity. But the question of the extent of decentralization required remains to be solved. As there is no definition of "sufficiently decentralized" proposed, nor is there, like in the US, any proposal of allowing a grace period for DAOs to mature to any given level of "sufficient decentralization" [29], such will likely have to evolve through regulatory technical standards set by the EU financial regulators. Combined, the typology suggests overlapping assumptions open for problematization [30].

This is further exacerbated by DAOs frequently operating across multiple jurisdictions with different views on decentralization, resulting in the matter becoming a topic of strategic importance as the uncertainty blocks investments, which impacts the competing growth and innovation objectives mentioned earlier.

## 2.3 Arriving at a Working Definition for Decentralization

The notion of "decentralization" has its origins in political science and, in the present time, generally refers to the dispersion or distribution of functions and powers. Without an understanding of the powers of different stakeholders, where and how they exercise their powers, and to whom and how they are accountable, it is difficult to understand whether decentralization is taking place [31].

The concept of decentralization has been applied mainly within the government of nation-states and political science [32], administration [33], fiscal area [34], and environment [35], but also across a diverse range of disciplines, such as complex systems engineering [36], space safety engineering [37], cybernetics [38], management science [39], economics around principal agents theory [40], finance [15], law and technology [41], crypto-economic systems [9] and more.

Within the nascent literature on crypto, the most applied definition of decentralization was proposed by Ethereum co-founder Vitalik Buterin with the introduction of the term "DAO" in 2013 [25].

Here, decentralization is presented as a response to the latent issues of centralized systems, to which decentralized systems can introduce fault tolerance and deter attacks or collusion. In a later publication [42], Buterin suggested that decentralization be viewed across several dimensions: (1) An architectural dimension as in how many computers the system is made up of; (2) a political dimension as in how many controls those computers; and (3) a logical dimension as in how the interface and data structures add up.

Some scholars and practitioners suggest that decentralization is a misleading term, as it has a slightly negative connotation, and no large-scale social, economic or political institution can be fully decentralized and automated without human intervention. Decentralization is then considered more specific to an activity, not to an organization design dimension; instead, we might consider using collaborative models [43].

It follows that measuring decentralization is complicated; "*A true assessment of the degree of decentralization in (a country) can be made only if a comprehensive approach is adopted, and rather than trying to simplify the syndrome of characteristics into the single dimension of autonomy, interrelationships of various dimensions of decentralization are taken into account*" [44], [45].

> We propose that "sufficient decentralization" is defined as a verifiable state, where (1) the design of the DAO is collusion resistant and based on long-term equilibrium; (2) its governance processes have unrestricted and transparent access.

## 3 Methodology

This article follows an inductive approach to framework development [46]. We chose thematic analysis as a method to reflect and unravel the surface of the "reality" of DAO decentralization [47] through interviews and literature review. We analyzed the data in six phases: (1) familiarize yourself with the data, (2) generate initial codes, (3) search for themes, (4) review themes, (5) define and name themes, and (6) produce the report.

We chose an explorative, qualitative research approach to identify the relevant dimensions of decentralization in a DAO. We conducted semi-structured, open-ended expert interviews to identify possible themes to supplement literature review findings.

Potential interviewees were approached through contacts from ongoing token engineering projects. We conducted eight interviews with experienced DAO experts and stakeholders (Table 1), each lasting 45–60 minutes.

At the beginning of each interview, we ensured proper consent and confidentiality. We used an interview guide [48] with 10 open questions probing the interviewees' perspectives on aspects of the structural elements of a DAO (decentralized, autonomous, organization) and additional dimensions for assessing decentralization specifically. Interviews were recorded and transcribed, amounting to 82 pages of transcripts and notes.

**Table 1.** Overview of Interviewees

| DAO Expert | Expert role | DAO experience |
|---|---|---|
| E1 | Complex Systems Architect and Designer | 6 years |
| E2 | Cryptoeconomist, token engineer, ecosystem designer | 4 years |
| E3 | Engineer, Data Scientist, DAO advisor | 5 years |
| E4 | Founder, DAO ecosystem tooling | 4 years |
| E5 | Serial entrepreneur, Co-founder misc DAOs | 8 years |
| E6 | Lawyer, Specialist in DLT/Blockchain projects | 5 years |
| E7 | Lawyer, Crypto Asset Specialist / DeFi legal expert | 5 years |
| E8 | Lawyer, DeFi specialist, National regulatory body | 5 years |

Although mainly conducted through one-to-one interviews in search of the "decentralization surface" of DAOs and with unclear requirements from the outset, our search process matches elements of a design science research (DSR) method [54], where the artifact design process informed an iterative process with stakeholders, leading to the final result. Our approach is summarized in Figure 2:

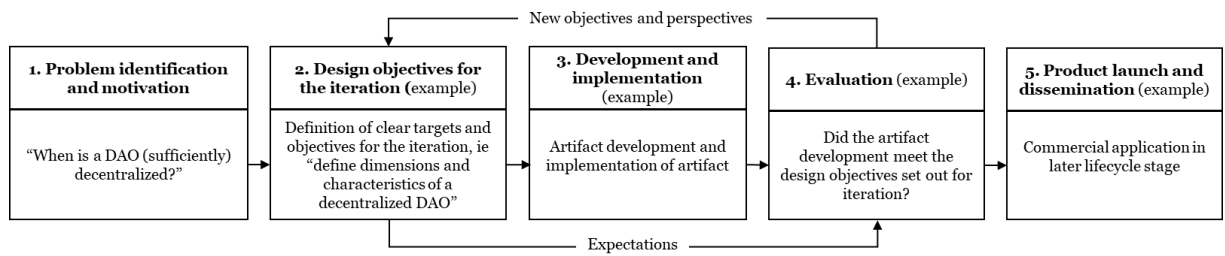

**Figure 2.** Our search process outline

After (1) reviewing transcripts and notes from interviews, we (2) extracted dimensions of decentralization and aligned them to the literature on DAOs and DeFi manually. The unit of analysis was the practices conducted by DAO communities, the subsystems used to perform these, and the technical infrastructure supporting them. All three authors were involved in the data analysis. As two authors were involved in the data collection, the third author maintained distance and acted as a devil's advocate to ensure the analysis remained objective and independent of our preconceptions and the interviewees' views [49].

As each expert had their own practical experience from working with DAOs, we first conducted a within-case analysis to gain familiarity with the data and generate a preliminary theory; then, we examined the data for cross-case patterns [50]. The coding procedure comprised several rounds of analysis and refinements of the codes. The topic of decentralization is multi-dimensional and complicated, having to determine the primary angle of analysis either by business subsystem, policy, or technical architectural dimension. During this procedure, we gradually moved from an inductive to an abductive approach [49], using labels to categorize the interviewee-specific language and grouping similar ones.

Our data sampling strategy remained open to new theoretical insights on what constitutes decentralization [51]. In (3) the search for themes, we clustered initial 52 first-order concepts across 7 DAO subsystems, 4 policy dimensions, and 4 technical architectural layers, further (4)(5) synthesizing these into 15 second-order themes across 5 aggregate dimensions. As we analyzed the data and generated theoretical concepts, we cross-referenced our findings with the extant literature in an iterative process to align our findings.

Our literature review followed a "light approach" [48], where we developed the research protocol, defined – and refined – the research question, and added criteria for DAO research while focusing mainly on decentralization and acknowledging related characteristics to autonomy and organization. The DAO subsystems were identified using a DAO reference model [52]. Still, as the framework should satisfy regulatory and supervisory expectations of a risk-based approach, we also investigated a technical reference model proposed by regulators [53].

Once we had derived the first-order concepts, second-order themes, and aggregate dimensions, we built the data structure as appears in Figures 3a and 3b below.

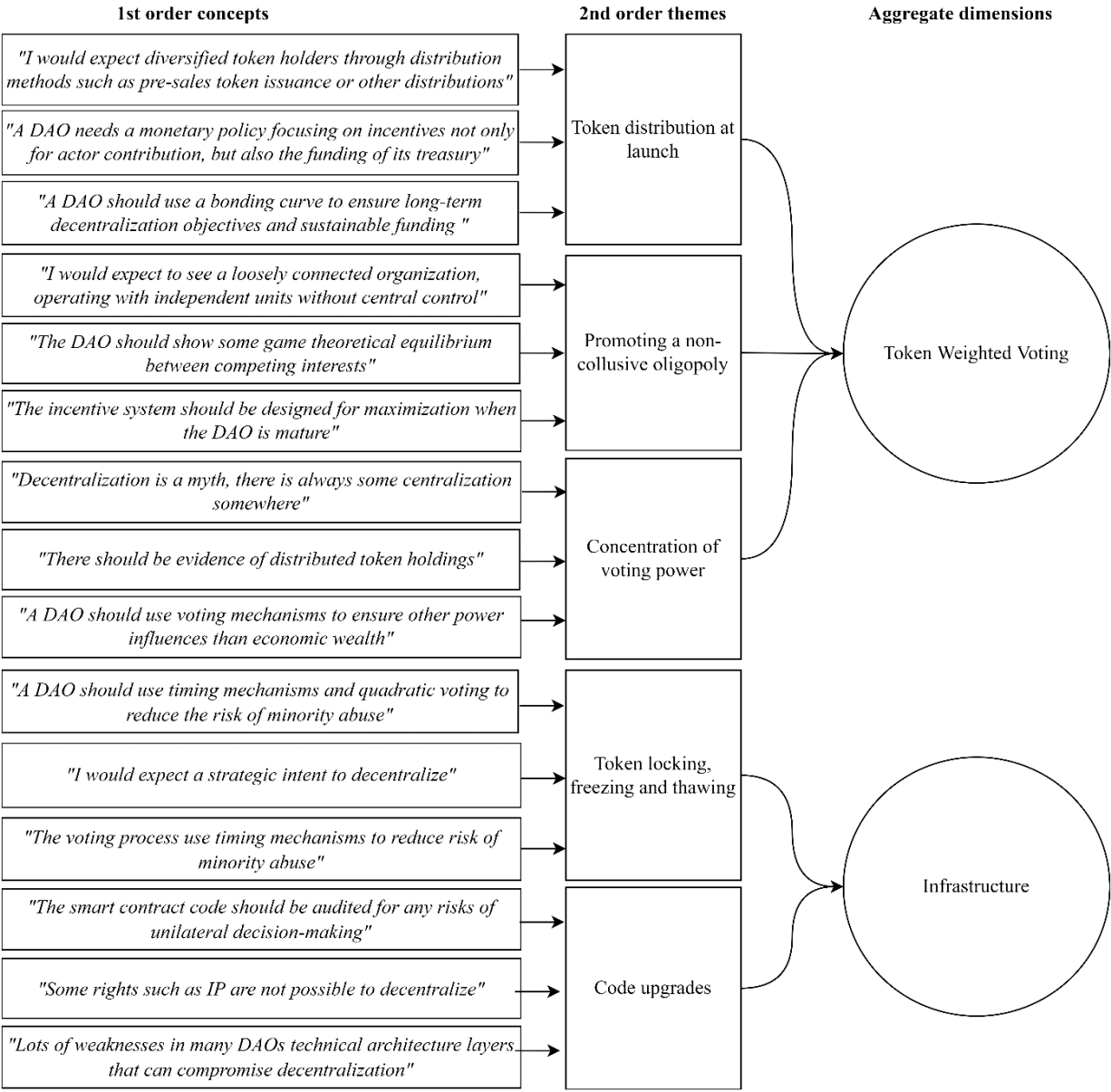

**Figure 3a.** Coding of data to themes (1 of 2)

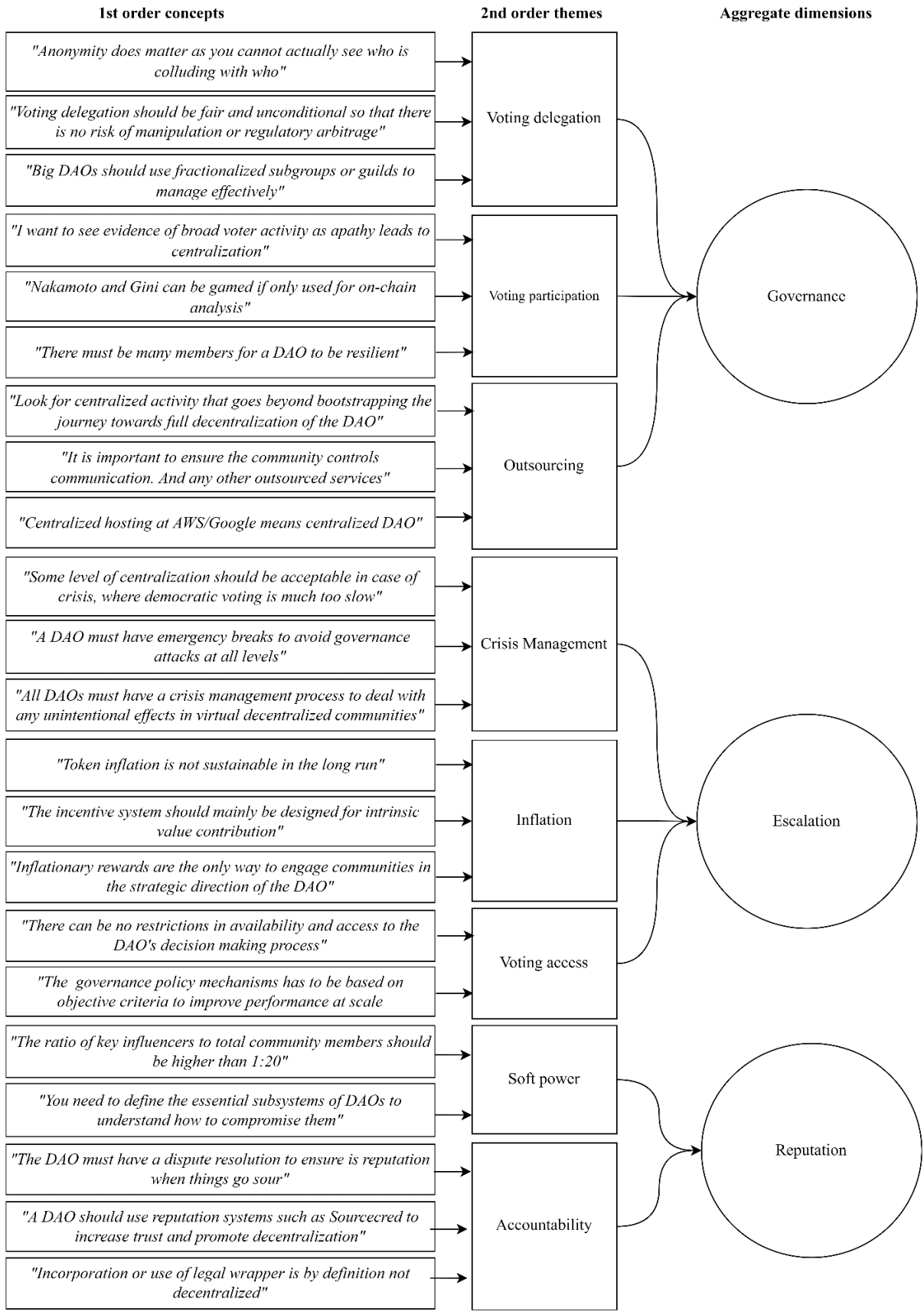

**Figure 3b.** Coding of data to themes (2 of 2)

The artifact was evaluated ex-ante by a representative from a regulator to ensure a level of alignment to regulatory expectations of the framework artifact.

# 4 Introducing "TIGER" Assessment Framework

The proposed artifact comprises a generalized DAO score-card evaluation framework. The framework facilitates a directional analysis of critical DAO components from a systems perspective, where compromising one subsystem may compromise the entire system [9], [43].

In the output component, we leverage traditional supervisory methods [55] and aim to score and consolidate each characteristic to generate an assessment score for each critical dimension that may affect the entire DAO level of decentralization if compromised. The central assessment approach is to which extent, on each dimension and its characteristics, we observe evidence of independent groups of agents operating under mandates without any centralized element of control.

The assessment is designed for point-in-time. Thus, no "safe harbor" assessment component is included, which could be relevant depending on the specifics of the DAO in question. We have, however, aimed to integrate strategic intent to allow a "grace period" to impact the scores. The actual application of scores requires some calibration and further consultations across DAOs and jurisdictions to evolve into a regulatory technical standard.

## 4.1 A Taxonomy of Agents in a DAO

Permissionless blockchains are essentially a vast network of databases maintaining a shared space. Transactions are batched and circulated with the network in the form of blocks which, once accepted by the network, amend the database with the most recent balance assigned to the known addresses. Maintaining a distributed database of transactions in this fashion introduces a high level of integrity. Still, it necessitates the encryption of user identities, as anyone with access to the database would otherwise be able to view the accounts balances of the individuals using the network.

Permissionless blockchains solve this issue with *private-key infrastructure* (PKI), in which a private/public key pair is used to generate any number of addresses. Traditional PKI is pseudonymous, as the user's identity is encrypted, but still predisposed to simple heuristic address clustering of transaction patterns [52]. As such, blockchain technology presents a fascinating paradox: Pseudonymous identities are essential in protecting user privacy but, at the same time, offer a design challenge for DAOs. Yet, the replicated nature of the database means that pseudonymous transaction data is available perpetually, enabling stakeholders to access the full transaction history for an address. Different agent definitions are shown in Table 2.

**Table 2.** Agent definitions

| Agent type | Description | Sample of Evidence |
|---|---|---|
| Verifiably Independent Agent (VIA) | A publicly identifiable token holder (maybe with a sizeable reputational interest in maintaining the integrity of their address) with a long and repeated history of participation in governance and a public presence in the associated communities. | Proof of (real or pseudonymous) identification measures across multiple governance discussions and social media sites, a discernible asset trail, and/or identification standard tokens (Ethereum naming service) |
| Presumably Independent Agent (PIA) | A token holder with a presumed vested interest in a sound governance process and | An address with a transaction history indicating repeated and non-automated use on a near daily basis, coupled with interactions in other DAOs and a discernible transaction pattern. |
| Unidentifiable Agent (UIA) | All addresses not operated by a PIA or a VIA. | Addresses with indications of automation and repetitive transaction patterns or clusters. |

## 4.2 The TIGER Assessment Questionnaire

After several iterations and pattern analysis, the conceptual artifact was optimized and consolidated to contain 15 characteristics with suggested questions and quantifiers for assessment as shown in Table 3. We summarize the requirements [56] in five general categories of DAO subsystems (items with grey background in column 1 of Table 3) based on expert input and literature [52]: Token Weighted Voting; Infrastructure; Governance, Escalation, and Reputation ("TIGER").

**Table 3.** TIGER Assessment Questionnaire

| Topical Analysis | | Variables |
|---|---|---|
| Category | Question | Quantifier |
| **Token Weighted Voting and Incentives** | | |
| Token distribution at launch | Did the team conduct a "fair" token launch designed to balance incentives for further decentralization with requirements for long-term funding and investor returns? | Percentage of units allocated to addresses associated with insiders, including core-team members, advisors, investors, early collaborators, and service providers. |
| Promoting a non-collusive oligopoly | Does the DAO algorithmically incentivize multilateral participation by rewarding non-colluding groups of agents for strategic participation? | Percentage of units allocated to clearly differentiated stakeholder groups indicated by a misalignment in assumed preferences |
| Concentration of voting power | How distributed are governance tokens amongst active/passive stakeholders? | Number of VIAs required to mount >51% of voting power in majority voting schemes? |
| **Infrastructure** | | |
| Token locking, freezing, and thawing. | Does the token contract code include the ability for any set of stakeholders to lock, move, freeze, and thaw token balances on some or all addresses? | Number of VIAs required to freeze token balances in all or some addresses. |
| Code upgrades | Is there evidence of the possibility of enforcing unilateral decision-making in the code that may compromise decentralization? While most code upgrades will preserve address, state, and balance, any ability to change smart contract code will impose significant security risks to the DAO and its stakeholders. | The number of agents of any type required to effectively implement a proposal or other non-specified changes to the smart contract code. Code changes or upgrades may be implemented either following official voting sessions or unilaterally. |
| Access | To what extent is access to decision-making through voting or other means accessible to external parties or contributors in a meaningful and unrestricted way? | Mixed assessment relating to quorum and timing: (1) How many verifiably independent agents does it take to produce a positive voting outcome for a "general" Improvement Proposal (Nakamoto co-efficient for governance), and (2) Does the voting process allow proper time and access for token holders to vote on any topic? |

| Topical Analysis | | Variables |
|---|---|---|
| Category | Question | Quantifier |
| **Governance** | | |
| Voting delegation | Is any voting delegation fair and unconditional so there is no risk of manipulating reported delegation? | How many VIAs with clearly distinctive preference profiles are presently available for delegation |
| Voting participation | Is there evidence of broad voter activity? | Percentage of token float with active participation in governance |
| Bootstrapping | Is there any centralized activity that goes beyond bootstrapping the journey toward full decentralization of the DAO? | *Qualitative assessment*: Is there evidence of centralized control measures that are not required for the long-term health of a decentralized DAO? |
| **Escalation** | | |
| Crisis management | Does the constitution or policies include crisis management and dispute resolution mechanisms? | Percentage of tokens required to enact crisis management decision-making |
| Inflation | What is the distribution between token inflation accruing to user A. External (oligopolistic) incentives for non-colluding VIAs (LPs, open-source developers, etc.) and user B. Insider VIAs such as investors, founders, early stakeholders, etc.? | The percentage split user A/ user B. |
| Voting access | Are there any restrictions on availability and access to the DAO's decision-making process? | Mixed assessment relating to quorum and timing: (1) How many VIAs do it take to produce a positive voting outcome for a "general" Improvement Proposal, and (2) if the voting process allows proper time and access for token holders to vote on any topic. |
| **Reputation** | | |
| Soft power | Is there evidence of co-optation or informal manipulation? | *Qualitative assessment*: Past evidence or forward-looking assessment of how many known high-profile agents can theoretically swing a vote |
| Responsibility alignment | Does the DAO code or applicable norms introduce the notion of accountability for decision-makers in a fashion that appears symmetrical to the power and responsibility vested in decision-makers? | *Qualitative assessment*: No evidence of asymmetry between responsibility and accountability, for instance, unjust overruling or veto. |
| Accountability | Are measures for conflict and reputation management implemented? | *Qualitative assessment*: Evidence of dispute resolution measures to mitigate centralized attack vectors around reputation |

### 4.2.1 Token-weighted Voting and Incentives

The assessment of this dimension includes:

- Analysis of whether the tokens are fairly distributed among the community, founders, and collaborators while also locking token liquidity for the future funding of the DAO's activities. Fair launch considerations include considerations over the pricing of the token across the issuance period(s). Essentially the assessment is a determination of whether the DAO's monetary policy is fair and whether anyone, including the core team, is benefiting unfairly compared to the DAO community long term.
- When assessing whether the DAO incentivizes multilateral participation by allocating tokens to clearly differentiated stakeholder groups, it is important to notice that some collaboration and common focus are to be expected. In addition to quantifying units allocated to independent groups, the assessor could also look for signals: Is there any tangible evidence of cartel's? Is it reasonable to assume that token holders are colluding unfairly? Are big investors talking to the founders and asking them what to vote for, or the other way around?
- The concentration of voting power would include a Nakamoto-coefficient analysis of on-chain and off-chain voting history. The Nakamoto coefficient is a simple, quantitative measure of a system's decentralization [57], [58]. The coefficient is based on the Gini coefficient and calculated based on the number of critical subsystems in a system and how many entities one would need to compromise to control each subsystem.

### 4.2.2 Infrastructure

The assessment includes:

- Analysis of how the DAO limits large token holders (so-called whales) from having outsized influence. Some DAOs introduce the notion of time-locked voting. This allows token holders to increase the weight of their vote by locking their shares for a certain amount of time after voting has ended, trading the opportunity cost for increased voting power. Freeze and thaw measures may also be applied to the benefit of late-joiners and/or to reduce whale influence.
- Analysis of centralization of control that is not automated in a sufficiently decentralized manner, which includes an assessment of the degree of autonomy in software vs. human centrality but also a view of any single point(s) of failure or single point(s) of control concerns.
- Access is assessed both to quorum and timing, assessing how many VIAs it takes to produce a positive voting outcome for a "general" Improvement Proposal, which we could label as the Nakamoto co-efficient for governance, and second, whether the voting process allows proper time and access for token holders to vote on any topic or if (unfair) restrictions apply.

### 4.2.3 Governance

Assessment of governance processes is critical to determine whether there are possible centralized attack vectors in a DAO:

- Voting delegation, sometimes referred to as liquid democracy, shares the core principles of political democracy. In this case, a DAO assigns specialists to participate in an electorate with the power to make decisions on behalf of DAO members. This increases centralization, on the other hand, it may improve the quality of decision-making as in the traditional world's representative democracies. In some cases, voting delegation may constitute manipulative and/or regulatory arbitrage through conditional delegation, so the assessment should review delegation mandates to ensure the delegated mandate is not an attempt to arbitrage. The analysis can range from a simple count of the number of individual components in the DAO network and the relative size of these to more advanced network analysis and statistical tests, where a DAO uses more advanced voting delegation.
- From a narrow perspective, the assessment of voting participation analyses voter turnout participation in collective decision-making, which is a dynamic metric that may affect the security of any plutocratic governance system. Simple token-weighted voting may risk the undue influence of "whales" (large token holders). Balanced techniques adopted by DAOs include sociocracy, where decisions are made by consent, not by consensus. Quadratic voting and other alternative voting mechanisms, such as holographic consensus or multi-signature wallet (multi-Sig), are also gaining traction across DAOs. The assessment may also include a fairness assessment of the voting process, where DAOs sometimes use timing mechanisms to reduce the risk of minority abuse. This process tackles the risk of majority voters gaining an advantage over minority voters; the downside is that the voting process becomes exceptionally long. Another method to ensure a fair voting process is "conviction voting," which is based on the community's aggregated preference and uses time as a utility to strengthen "conviction" to one's vote. A third example includes express voting that may encapsulate intensity or broader community support and thereby reduce the costs of democratic coordination.

- Sometimes, DAOs establish a foundation to own rights that can not easily be decentralized. Although this implies a centrally controlled activity, it should be viewed in context and be considered acceptable if the purpose of the centralized effort is only to bootstrap the journey towards decentralization. Outsourcing also includes software deployment strategy and hosting policy, where, according to statista.com [59], more than 64% of the world's cloud market is currently controlled by three dominant vendors (AWS, Google, and MSFT), who therefore likely host most of the blockchain/Web3 infrastructure that exists, including full nodes, validator nodes, and middleware. This is potentially a significant attack vector for censorship and centralized control.

### 4.2.4 Escalation

Consideration of the following issues helps in assessing escalation:

- A DAO is only as decentralized as its crisis mode allows. Hence, the assessment should investigate how control measures can be centralized in any crisis. A crisis should be defined through stress testing of the DAO business system and financial and technical resilience. Crisis mitigation and contingency measures should preferably be specified in the DAO constitution or policies for events that can impact the long-term sustainability of the DAO. Some centralization is expected to deal effectively with crisis containment, where fluid democracy may not always be the most efficient. Still, the assessment should determine the extent to which such centralization is subject to democratic control.
- An inflationary token model adds new tokens to the market over time, often through a schedule or as mining rewards or for specific contributions. For the determination of decentralization, the critical assessment point is that any value associated with inflation or deflation benefits all token holders fairly, not for the benefit of non-collaborative agents for any strategic or other participation.
- Availability and access should be equal to all, so any restrictions in access to the DAO, including its decision-making process, may suggest a level of centralized control. The assessment would include a Nakamoto coefficient analysis for both on- and off-chain activities around voter activity and token holdings and a review of voting policies.

### 4.2.5 Reputation

For assessment of reputation, the following considerations are suggested:

- Soft power through co-optation or informal manipulation is an everyday phenomenon in politics. In DAO communities that allows actors to engage pseudo- or anonymously, it is critical to assess that these features are not used manipulatively. Again, the analysis may potentially involve sophisticated network and statistical analysis.
- DAOs cannot act outside their rules, but because their smart contracts may contain errors or unforeseen events may occur, rule change mechanisms are necessary for resilience purposes. On the other hand, fully decentralized DAOs must also acknowledge their delegated mandates, with accountability following delegated responsibility.
- Increasingly, DAOs implement dispute resolution mechanisms or use dispute resolution services from emerging online third-party decentralized dispute resolution service providers. Other measures, such as implementing tools like Sourcecred [60] to create trust in the community, or slashing to penalize unwanted behavior or dishonest validation, are similar mechanisms of democratic control designed to incentivize network participation.

## 5 Evaluation

The artifact evaluation was conducted two-fold; First, we field-tested the general concept with a DeFi expert from an EU-based supervisory authority. Second, we applied the TIGER framework to a prominent DAO using publicly available sources.

The field-test evaluation emphasized a pragmatic approach favoring comprehensive coverage of topics of regulatory concern rather than the collection of quantitative data. The introduction of partial compromisation having a full impact on the overall assessment result was deemed justifiable but raised several questions, including (1) how to deal with the lack of a grace period in the current implementation of the recently released MiCA package and (2) how to create a level-playing field for "institutional DeFi" (where traditional, currently regulated financial institutions offer decentralized financial products operated by DAOs).

In the remainder of this section, we present a sample evaluation of a DAO as a reference guide to how regulators or industry participants may approach the discretionary application of the TIGER framework.

We use the Compound protocol and its associated governance processes for the sample evaluation. It is important to note that the sample application provided here serves only as a reference guide due to the lack of access and transparency for internal data. While DAO governance primarily happens in public fora, a regulatory authority would arguably have access to a wealth of quantitative and qualitative data provided and collected by the counterparty and its partners.

While this level of access is not attainable in the academic context due to privacy regulations, the level of public governance data available is sufficient in providing a cursory reference application of the framework. Further, if a DAO is already decentralized before enforceable regulation is agreed upon, a regulator/supervisor will need to rely on the same publicly available information we access here. The Compound protocol offers an interesting entry point to the evaluation of the TIGER framework, as the protocol team was amongst the first to issue a governance token (COMP) and the adjacent infrastructure, which led to the present generation of DAO governance.

While stablecoin issuer MakerDAO had already issued their governance token (MKR) years prior, the Compound team was amongst the first to explicitly link the issuance of the token with the usage of the protocol in a bid to incentivize liquidity provisioning. This sparked a period of rapid escalation, commonly referred to by industry observers as "DeFi Summer," in the 3rd Quarter of 2020 as the major decentralized exchange Uniswap (UNI) immediately followed suit in a bid to defend market share against aggressive attempts at siphoning liquidity by the rapidly emerging competitor "SushiSwap" (SUSHI). The ensuing period saw waves of governance tokens enter the market, mimicking the previous ICO frenzy [61].

### 5.1 Introducing the Compound DAO

Compound [62] is an on-chain market for peer-to-peer lending, enabling users to collateralize and borrow against a selection of 18 assets. At the time of writing, the protocol manages ~€3.7bn in collateral assets deposited by ~300 000 depositors, of which ~9000 users have taken out an aggregate of ~€895m in outstanding debt against their deposits.

Protocol decision-making is governed by token-holders utilizing the token (COMP) within the governance contract. The Compound Governance process involves submitting pre-deployed code changes to risk management and asset modules above, which stakeholders can then inspect and vote for or against implementing in binary voting sessions. Proposals are generally used to implement system parameter modifications, but proposals for adding new markets or entirely new features are occasionally implemented as well.

Further in this section, we present a cursory application of the TIGER framework, utilizing a score-card methodology in which we assign a score between 1–5 for each dimension. While there are clearly identifiable areas of improvement, we assess that the Compound DAO is *sufficiently* decentralized when we factor in the protocol age. Over time, we expect a gradually increasing decentralization as the protocol matures and increasingly larger private and institutional stakeholders join the DAO.

The overall score of our assessment is 3.8 on a scale of 5, split on each aggregate dimension as appears in Figure 5, with no critical dimension failing. A detailed assessment follows below.

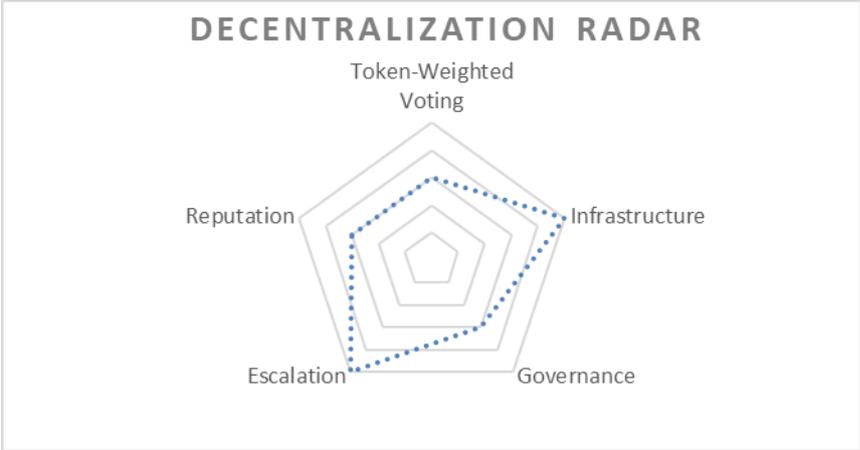

**Figure 4.** Compound decentralization radar

## 5.2 COMP Token Weighted Voting Distribution

The COMP token has a max supply of 10m units, of which 7.15m is in circulation at the time of writing. The COMP supply has a daily inflation rate, currently set at 1139 COMP daily, distributed across market participants (Table 4), alongside a 4-year vesting period for insider shareholders ending in June 2024.

**Table 4.** COMP allocation to stakeholder groups[†]

| Stakeholder Groups | COMP Allocation | Percentage of Total Supply |
|---|---:|---:|
| Shareholders of Compound Labs, Inc. | 2 396 307 | 23.96% |
| Founders & team | 2 226 037 | 22.26% |
| Future team members | 372 707 | 3.73% |
| Users | 4 229 949 | 42.30% |
| Community Allocation | 775 000 | 7.75% |

As evident, the COMP tokens allocated to shareholders in Compound Labs, Inc. Founders and team members (present and future team members) comprise a narrow minority share of 49.95% of the total token supply, assuming that the recipients retain all tokens after vesting.

While the narrow minority does not technically produce a concentration of voting power in the hands of stakeholders with presumed shared interests, it should be noted that in the theoretical event of a highly contentious issue between insiders and (external) community members, challengers would need to mount 50.05% of the token float to push through a decision, which is deemed unlikely.

---

[†] https://messari.io/asset/compound/profile/supply-schedule

Yet, the distribution of tokens amongst smart contracts and agent types [63] is such that, at present, only a few VIAs retain an adequate amount to mount a hostile proposal process. On this basis, we assign a passing score of 3 out of 5, informed by the relative concentration of votes.

### 5.3 COMP Infrastructure Assessment

The Compound team has implemented a well-reasoned and simple user interface for the governance process, enabling non-technical users to participate in the governance process.

The Compound Governor and Timelock methods require the deployment of code with the proposal submission. From proposal submission through voting and the mandatory two-day delay following a successful vote, the governance process implements a full week period for any decision made by DAO stakeholders.

In contrast to the frequently used option of using the popular tool Snapshot [64] to collect votes through signatures, this methodology mitigates the need for a single or multi-signer solution which can be required to implement the results of the vote when using Snapshot. Instead, approved proposals are immediately implemented by the contract once they pass. While this methodology has previously imposed costs on voters due to the high execution fees on the Ethereum blockchain, the team has implemented the casting and delegation of votes by offline signatures [65], mitigating voter apathy and improving accessibility of governance participation. Delegation functionality is implemented in the COMP token contract and delegates the voting power for the tokens from one address to another. Users interested in delegating voting power to multiple delegates can split tokens over multiple accounts and delegate to multiple delegates. The COMP token smart contract does not allow freezing addresses, manipulating balances, or upgrading the contract code through upgradeable "proxy contracts."

On this basis, we assess that the Compound governance model and the associated smart contract infrastructure are sufficiently decentralized, yielding a 5/5 score.

### 5.4 COMP Governance Dynamics

The Compound governance model utilizes delegation strategies, through which token holders can delegate voting power to active participants. To create a proposal, an address must hold in excess of 25 000 COMP (€1.5m) or lock 100 COMP (€6000) to create an "autonomous proposal," which can become ratified if delegated an excess of 25 000 COMP.

Governance proposals are time locked in review for three days, after which voting is initiated for an ensuing three-day period. Proposals gathering a majority of votes with a lower threshold of 400 000 COMP votes are queued for implementation for two days.

The governance of Compound is primarily in the custody of the delegate VIAs, retaining an aggregate of 92.6% of voting power with 2 377 404 COMP tokens in delegation. Of the top 60 delegates, accounting for 99.9% of the total voting weight, there is no additional delegation, so it is fair to assume the said VIAs also control these tokens.

The VIA delegates yield decisive authority over the Compound protocol, for which approximately 70% of the 36 proposals decided upon in 2022 (including failed and canceled votes) were decided by less than ten delegates wielding a clear majority. So far, in 2022, on average, ~600 000 COMP was active in each proposal, again mainly controlled by VIAs.

Through the lifetime of the DAO, 113 proposals have been voted upon, averaging 2.3 per month. The average voter turnout has increased slightly over time to 66 participating addresses per proposal in 2022, up from 56 addresses per proposal in 2020, the first year of operation [66].

Based on this assessment, it appears evident that while Compound governance is managed by a relatively small subset of VIAs with, in most cases, presumed identical preferences, said stakeholders would be unlikely to mount a hostile proposal against users, given the token distribution.

On this basis, we assign a passing score of 3 out of 5, informed by the relative concentration of votes.

## 5.5 COMP Escalation and Crisis Management

The Compound governance system uses timelock to introduce sufficient time for careful review of the proposal code before implementation. The community implemented an automated "Proposal Threshold Alert" as an early indicator of potential governance attacks. The alert informs the community if a wallet has accrued sufficient COMP to meet governance thresholds. Further, the Compound Comptroller contract includes elements of a crisis management mechanism with a pause guardian. Compound Labs previously controlled this, but since 2021 transferred it to a community multi-Sig wallet created by community members, where a small group of 4–6 stakeholders, chosen by the community, can pause Mint, Borrow, Transfer, and Liquidate functions. In our understanding, this does not constitute a complete "emergency shutdown" mechanism, so we assess that the multi-Sig does not provide full crisis management capability.

The lack of any special escalatory privileges awarded to early stakeholders became evident early in the life of the protocol when a bug in a proposal placed 280 000 COMP tokens at risk of emission to liquidity providers. While the Compound team removed the ability for users to claim these tokens through the interface, this did not stop users from simply interacting directly with the smart contracts.

In what appears to be a somewhat misguided attempt to return the tokens to the protocol, the founder of Compound Labs, Robert Leshner, threatened to collect information on non-cooperative stakeholders to inform the US tax authorities [67]. While these attempts were ridiculed by the community members, the case resembles the user B situation in Figure 1 above. It provides an example of how all stakeholders, regardless of their seniority in the community, cannot influence decisions governed through smart contracts.

Based on the lack of discriminatory privileges awarded to key stakeholders, outside of the ability to amend the contract web interface, we assess that the Compound DAO is sufficiently decentralized on this dimension, yielding a score of 5/5.

## 5.6 COMP Reputation and the Impact of Soft Power on Decision-Making Processes

Compound governance primarily occurs in designated online fora, where governance participants pitch and discuss proposals before developing and deploying a proposal code. Discussions are generally cross posted on social media [68] with parallel discussions occasionally led on chat servers [69]. On average, new posts are submitted daily to bi-weekly, indicating a moderate to high activity level.

By cross-referencing with data from LinkedIn [70], we note that the official organization appears to employ 19 employees with titles indicating a commercial relationship with Compound Labs Inc. We did not find evidence of any inordinate influence in proposal submissions by these employees. However, the picture is different when we assess the influence of large vs. small token holders in what we presume is the primary governance forum [71] for pre-proposal discussions: Out of a total of 113 proposals to date, 97 are included in the pre-proposal discussion. Of these, at least 53 posts have been authored by individuals in founding roles or with clear connections to the founding team or major token holders. Of these 53 posts, 32 were authored by the service provider Gauntlet [72], a firm specializing in financial modeling, which previously completed a market risk assessment report on Compound [73]. Gauntlet is identified as the controller of the fourth biggest delegate address, yielding 118 494 COMP at the time of writing this article. While Gauntlet is a frequent and active participant in Compound governance, the primary emphasis is on topics clearly related to risk management or the addition of new assets to the platform and does not appear manipulative.

There appears to be no dispute resolution mechanism. In the Compound chat forum on Discord; this has been debated, with some community members objecting to any dispute resolution mechanism and others firmly in support. The topic has not been subject to a formal vote. On this basis, we assign a score of 3 out of 5 on this dimension.

## 6 Discussion

In this article, we propose an information system (IS) focused conceptual artifact based on a review of the literature, combined with expert insights from a group of industry stakeholders and experts. The artifact demonstrates the feasibility of structured assessment methods of the level of DAO decentralization both on-chain and off-chain, mapped to generalized, critical processes of DAOs. We address the research question: "When is a DAO (sufficiently) decentralized?"

In analyzing whether a DAO is sufficiently decentralized, we might expect some quantified evidence of chaos, swarm, and/or a self-organized, distributed, decentralized community, as opposed to an ordered, strong organization with centralized command and control that characterizes the traditional organization.

Hence, the critical focus of analysis is whether the DAO stakeholders or "actors" are empowered with delegated authority and whether they operate sufficiently independently of each other and in their own self-interest in an uncoordinated and voluntary manner.

We propose that "sufficient decentralization" is defined as a verifiable state, where the design of the DAO (1) is collusion resistant and based on long-term equilibrium, and (2) its governance processes have unrestricted and transparent access.

From a regulatory perspective, an alternative approach could simply be to analyze (1) if the DAO is conducting a regulated activity, and if so, (2) if there is an accountable legal or physical person upon whom regulation can be enforced; if not, then DAO being sufficiently decentralized has to be acknowledged. In our view, such an approach is too simplistic and does not accept the fundamental premise that DLT/Blockchain is a transformative technology that will foster innovation and growth.

In terms of conciseness and robustness attributes of the assessment framework, the challenge lies in the complexity of decentralization as a concept. We avoid an extensive classification scheme that could lead to cognitive overload when assessing a given level of decentralization point in time while also defining enough dimensions and characteristics to clearly differentiate the objects of interest [55].

From a practical and theoretical perspective, it seems evident that no DAO can start decentralized, as any project must be initiated by a small core team, bootstrapping development until the project matures and attracts open-source contributors. However, as discussed, the European regulators did not play any particular emphasis on this critical point when agreeing on the final text of the MiCA regulation. Some US regulatory proposals suggest a safe harbor rule [25], proposing a grace period to allow a DAO to become sufficiently decentralized, thus introducing the concept of "gradual decentralization." In our proposed assessment framework, we acknowledge this by suggesting that the assessment includes a perspective on the mature DAO design, not just the point-in-time view.

We extrapolate our contributions into the following generalized propositions:

**P1**: The concept of technology-neutral regulation is challenged by DLT/Blockchain. DAOs exist and realize benefits through increasing degrees of decentralization. DAO legal design should therefore support the internal decentralization accomplished by the DAO so that a balance is achieved between external and internal decentralization [11], not the other way around. When regulators in the coming years design technical requirements for the supervision of DAOs, they need to acknowledge this underlying premise and embrace that DLT/blockchain is a transformative technology that requires unique regulatory approaches.

**P2:** Regulators need to embrace the concept of a "grace period" for a DAO to achieve sufficient decentralization. The MiCA regulation did not include this, but it seems challenging to embrace DeFi and the concept of sufficient decentralization without it. We suggest an assessment approach where not only the point-in-time assessment is material to the decision of decentralization but also the design intent, thereby introducing a grace period from a risk-based perspective, allowing the EU to practically align crypto regulatory compliance to the safe harbor proposals from the US [25] and common sense.

**P3**: In the short term, for "Institutional DeFi," a level playing field needs to be developed by financial regulators and supervisors, including a "cut-off" strategy, with clear boundaries for acceptable centralized activity, to allow DLT/Blockchain-based businesses to develop properly, respecting the new technological feature regime. From a regulatory perspective, and in the words of MiCA, complete decentralization seems to require full automation. Still, when elements of human governance are introduced, it is difficult to think of complete decentralization as outlined in MiCA. Some automated features also become centralized through the front-end website hosting or other elements. Regulators must accept that a new playing field for DAOs will develop over the coming years.

**P4:** Regulatory practices around DAO decentralization will evolve across blockchains and business models, each with its own strengths and weaknesses regarding centralized attack vectors and regulatory importance. A risk-based approach to DAO supervision, where required, will therefore need to be developed with a holistic view of decentralization across political, technological, social, and economic dimensions, as well as across underlying technology infrastructures that behave very differently from a risk perspective. We foresee regulators will designate some blockchains to have more systemic risk than others.

**P5:** DLT/Blockchain will transform how regulators supervise and enforce the regulation. The number of DAOs grew by a factor of 8x in the past year [74]. With the increasing certainty on the regulation of crypto, the number of DAOs will likely continue to evolve, and the growth of the token economy and innovation of blockchain-based business models as well. Some sample DAO business models [76], [77] are listed in Appendix 1.

These developments pressure regulators to keep pace with developments in two dimensions: (1) Supervisors with a traditional finance focus will be challenged as their supervisory toolkits and skillsets become disconnected and obsolete. Regulators and supervisors must embrace the available and emerging investigative techniques to analyze DAO structures and processes in real-time, on- and off-chain; (2) A focus on automated and embedded supervision should be prioritized [75].

Our work contributes to practice by identifying criteria for DAOs, regulators, and supervisors to consider when assessing whether a DAO is "sufficiently decentralized," complementing the understanding beyond technical difficulties by taking a holistic view of DAOs as complex socio-technical systems.

Our findings contribute actionable insights to the information system literature by emphasizing how DLT and blockchain technologies may be assessed from a socio-technical perspective. We contribute to DAO communities and regulators with a pragmatic tool to understand to what extent an otherwise regulated activity may be considered sufficiently decentralized and thereby avoid significant and costly compliance requirements.

# 7 Conclusion

We investigate the topic of decentralization as it relates to DAOs, using a thematic analysis method to identify relevant patterns to assess whether sufficient decentralization is presented. Through the framework's design, we demonstrate the feasibility of implementing a structured method for the assessment.

We propose a definition of "sufficient decentralization" and incorporate the notion of a representative democracy via delegated mandate in the assessment framework. Still, it remains to be concluded what level of delegation and decentralization is acceptable under different regulatory regimes. Some regulators seem to suggest complete decentralization as the only acceptable level. However, complete decentralization in DAOs is challenging to grasp, as they are socio-technical constructs.

We design a generalized assessment framework with suggested quantifiers. Still, the application of all characteristics and levels of quantified assessment will likely vary, depending on the need for regulatory monitoring by jurisdiction. Hence, the framework design is flexible to accommodate change as regulatory practices evolve and regulatory technical standards become defined. We demonstrate the practical application of the framework artifact by assessing the level of decentralization of Compound, an algorithmic money market DAO operating on the Ethereum blockchain.

Our findings suggest that decentralization in DAOs is not a myth. Still, due to the technical features of blockchains, it can be complicated to investigate and assess the true level of DAO decentralization. Our contribution is a pragmatic framework that can guide aspiring DAOs, regulators, and supervisors to advance the decentralization agenda as the crypto and traditional economies increasingly overlap and integrate. We extrapolate the findings into five general propositions on the implications of decentralization on the supervision of regulated financial activity in crypto.

## Acknowledgments

The authors wish to thank the anonymous reviewers as well as Danny Dehghani, Jon Isaksen, Michael Zargham, Griff Green, Angela Kreitenweis, Nina Siedler, Marina Markezic, Kris Paruch, and Matthew Barlin for their valuable insights and feedback.

# Appendix 1 – Sample DAO Business Models

| Category * | Description |
|---|---|
| Media DAO | Media DAOs such as Mirror (https://mirror.xyz/) empower writers and make it possible to work alone or collaboratively to publish, crowdfund, and create auctions and editions of media projects or digital artwork through tokens. |
| DAO Operating system | DAO operating systems or "platforms" such as Aragon (https://aragon.org/) or DAOstack (https://daostack-1.gitbook.io/v1/) provide a complete software stack and infrastructure for building and running a DAO, including various apps for token management, voting, and finance. |
| Social DAO | The Social or Community DAO category covers a broad range of DAOs that focus more on social capital than financial capital; they include communities that evolve from group chats to co-working DAOs or just a meeting place. An example is Filmmaker DAO (https://www.filmmakerdao.com/), which coordinates filmmakers' efforts to enable more IP ownership. |
| Protocol DAO | Protocol DAOs were initially intended to transition power from a founder team into a broader community, finding new ways for projects to issue fungible tokens into the market. These DAOs now constitute the bulk of decentralized finance (DeFi) protocols, such as Aave (https://aave.com/), Uniswap (https://uniswap.org/), or MakerDAO (https://makerdao.com/) and typically with a transaction focus aiming to compete with traditional finance. |
| Collector DAO | Collector DAOs are the home of NFT art-focused DAOs, such as PleasrDAO (https://pleasr.org/) enable their community to share the cost of expensive assets and co-own digital art, in the case of PleasrDAO specializing in what the members determine are culturally significant art pieces, that are further fractionalized for trading on DeFi protocols such as Uniswap v3 NFT. |
| Investment DAO | Investment DAOs such as Seed Club DAO (https://www.seedclub.xyz/) enable their community to co-invest, build and accelerate digital communities, land, or other assets deemed relevant for an investment focus. |
| Impact DAO | Impact DAOs, such as Climate DAO (https://climatedao.xyz/), focus on sustainability and conservation agendas. They are frequently driven by activist communities collaborating with research institutions or having educational activities. |
| Service DAO | Service DAOs, such as BrightID (https://www.brightid.org/) support DAOs with all required infrastructure and operational services, for instance, token, governance, or operational services, including voting, recruitment, legal, risk management, community management, technology, treasury, or, in the case of BrightID, a decentralized digital identity DAO. |
| Grants DAO | Grant DAOs such as Gitcoin (https://gitcoin.co/) enable their communities to donate funds and vote through governance proposal rounds on how the distributed funding capital is allocated to various projects, typically focusing on digital common goods aligned with Ostrom principles and not for profit. |

\* Sources: "DAOs List - Messari." https://messari.io/governor/daos (accessed Jul. 24, 2022) and "Full-Time DAOs — Coopahtroopa." https://coopahtroopa.mirror.xyz/5vTIKBRzMpVAiNyc7CnABXjh3ToJrjQOnOdkwqvb3l8 (Accessed on Jul. 24, 2022).